\documentclass[12pt]{article}
\usepackage{mathpazo}
\usepackage[T1]{fontenc}
\usepackage[utf8]{inputenc}

\usepackage{graphicx}
 \usepackage{booktabs}

\usepackage{dcolumn}
\usepackage{bm}
\usepackage{braket}
\usepackage{amsmath}
\usepackage{relsize}
\usepackage{hyperref}                   	
\usepackage{lipsum}
\usepackage{microtype}                      
\usepackage{fancyhdr}
\usepackage{tabularx}
\usepackage{makecell}
\usepackage{picture} 
\hypersetup{								
	colorlinks=true,      		            
	linkcolor=blue,                        	
	filecolor=blue,                     	
	citecolor=blue,    	        			
	urlcolor=blue,                 	   		
	pdffitwindow=false,               	 	
	pdfstartview={FitH},               		
	pdfauthor={Wharton \& Price},           		
}

\usepackage{comment}

\usepackage{tikz}
\usetikzlibrary{shapes,decorations,arrows,calc,arrows.meta,fit,positioning}
\tikzset{
    -Latex,auto,node distance =1 cm and 1 cm,semithick,
    state/.style ={ellipse, draw, minimum width = 0.7 cm},
    point/.style = {circle, draw, inner sep=0.04cm,fill,node contents={}},
    bidirected/.style={Latex-Latex,dashed},
    el/.style = {inner sep=2pt, align=left, sloped}
}

\begin{document}

\title{W as the Edge of a Wedge: Bell Correlations via Constrained Colliders}\label{bell}
\author{Huw Price\thanks{Trinity College, Cambridge, UK; email \href{mailto:hp331@cam.ac.uk}{hp331@cam.ac.uk}.}
}
\date{\today}
\maketitle\thispagestyle{empty}

\begin{abstract}
In previous work with Ken Wharton, it was proposed that Bell correlations are a special sort of selection artefact, explained by a combination of (i) collider bias and (ii) a boundary constraint on the collider variable. This requires no direct causal influence outside lightcones, and may hence  offer a new way to reconcile Bell nonlocality and relativity. This piece outlines a new argument for the proposal. It explains how it is valid for a special class of (`W-shaped') Bell experiments involving delayed-choice entanglement swapping, and argues that it can be extended to the general (`V-shaped') case. (A detailed version of the argument is now available in \cite{PriceWharton24}.)
\end{abstract}

\maketitle

\section{Introduction}
Bell's Theorem \cite{Bell64,Myrvold21} appears to show that if certain quantum predictions are correct, measurement choices on one wing of an EPR-Bell experiment are not independent of outcomes on the other wing, even if the two systems are spacelike separated at the time of measurement. This failure of independence is Bell's famous `nonlocality'. The predictions concerned (so-called `Bell correlations') have now been confirmed in many experiments \cite{Hensen15,Giustina15,Shalm15}, though their implications remain disputed.  

In previous work \cite{PriceWharton23}, Ken Wharton and I have proposed a new explanation for Bell correlations, arguing that they can be regarded as a special sort of selection artefact. If correct, this proposal not only explains the origin of Bell nonlocality, but also renders it compatible with relativity. 

The present piece outlines a novel argument for the proposal. It is in three steps.
\begin{enumerate}
    \item We begin with a class of `W-shaped' Bell experiments, involving delayed-choice entanglement swapping (DCES). As several authors have pointed out, Bell correlations in such experiments do not require nonlocality, being explicable instead as selection artefacts. 
    \item We note that the situation changes if the outcome of the entanglement swapping measurement is \textit{constrained} to take a single value, in the way suggested by Horowitz and Maldacena to apply within black holes. In effect, this constraint converts a selection artefact into genuine nonlocality. (This is crucial to the Horowitz-Maldacena hypothesis, which is a proposal about how information escapes from black holes.) In this very unusual case, then, Bell nonlocality results from imposition of a boundary constraint on a correlation that would otherwise be a mere selection artefact. 
    \item This highly unusual W case then acts, appropriately, as the edge of a wedge. We argue that the same explanation of nonlocality works in ordinary `V-shaped' Bell experiments, where the relevant boundary constraint is supplied by ordinary experimental control of the initial state. The parallel has been missed, apparently, because the boundary constraint is ubiquitous in the V-shaped case, but exceptional in the W-shaped case. However, the structure of the explanation of nonlocality is the same in both cases.
\end{enumerate}  

 It is a common view that the Bell correlations exclude \textit{local realism,} but leave two options: reject \textit{locality} or reject \textit{realism} (see \cite{Maud14,Gomori23,Wise17} for discussion).  We stress that rejecting realism is not our strategy. Our aim is not to avoid nonlocality, but to explain it, in a relativity-friendly fashion.

\section{The argument}
\subsection[The path to V via W]{The path to V via W}
\begin{figure}[t]
\centering
\begin{tikzpicture}
    \node[draw,rectangle,minimum width=0.6cm,minimum height=0.5cm] (I1) at (3,-2.5) {\textsf{S$_1$}};
    \node[draw,rectangle,minimum width=0.6cm,minimum height=0.5cm] (I2) at (7,-2.5) {\textsf{S$_2$}};
    \node[draw,rectangle,minimum width=0.8cm,minimum height=0.5cm] (W) at (5,3) {\textsf{C}};
\node[] (W0) at (4.4,4) {\textbf{0}};
\node[] (W1) at (4.8,4) {1};
 \node[] (W2) at (5.2,4) {2};
 \node[] (W3) at (5.6,4) {3};
\node[] (Out) at (5,4.7) {$C$};

 \path [dashed,-] (W1) edge (W);
 \path [line width=1.5pt,-] (W0) edge (W);
\path [dashed,-] (W2) edge (W);
 \path [dashed,-] (W3) edge (W);

    \node[draw,rectangle,minimum width=0.6cm,minimum height=0.5cm] (Abox) at (1,0) {\textsf{A}};
  \node[] (Alabel) at (1,1.7) {$A$};
    \node[] (A0) at (0.6,1) {0};
    \node[] (A1) at (1.4,1) {1};
    \node[] (a) at (0.3,-1) {$a$};
    \node[draw,rectangle,minimum width=0.6cm,minimum height=0.5cm] (Bbox) at (9,0) {\textsf{B}};
      \node[] (Blabel) at (9,1.7) {$B$};
       \node[] (B0) at (8.6,1) {0};
    \node[] (B1) at (9.4,1) {1};
     \node[] (b) at (9.7,-1) {$b$};
       \node[] (bc) at (4.4,5.2) {\footnotesize{\textbf{Postselection}}};
       \path [<-] (bc) edge (W0);

   \path [dashed,-] (Abox) edge (A0);
   \path [dashed,-] (Abox) edge (A1);
    \path [->] (a) edge (Abox.south west);
    \path [dashed,-] (Bbox) edge (B0);
   \path [dashed,-] (Bbox) edge (B1);
   \path [->] (b) edge (Bbox.south east);
      \draw [color=blue,line width=1.5pt,-] (I1.north) -- (Abox.south east);
   \draw [color=blue,line width=1.5pt,-] (I2.north) -- (Bbox.south west);
      \draw [color=blue,line width=1.5pt,-] (I1.north) -- (W.south);
       \draw [color=blue,line width=1.5pt,-] (I2.north) -- (W.south);

  
\end{tikzpicture}
\caption{The W-shaped case with postselection for outcome $0$ at \textsf{C}} \label{fig:W}
\end{figure}
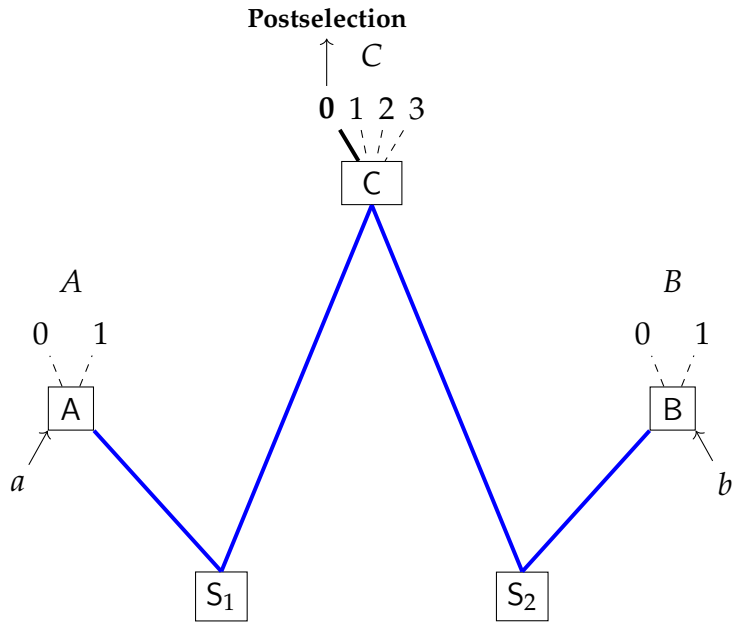

Our goal is to explain the Bell correlations in an ordinary two-particle V-shaped Bell  experiment of the kind shown in Figure~\ref{fig:V}. We begin with a different sort of Bell experiment: a four-particle W-shaped case involving delayed-choice entanglement swapping, as in Figure~\ref{fig:W}. (It is easy to choose versions of these experiments so that the two experiments exhibit precisely the same correlations between outcomes and settings, but this is not essential to our main point.)

W-shaped Bell experiments rely on \textit{postselection.} The procedure requires that we retain only the cases in which the entanglement swapping measurement \textsf{C} has one of four possible outcomes, and then look for  Bell correlations between the settings and outcomes  $\{a,b,A,B\}$ of the measurements \textsf{A} and \textsf{B}, in this subensemble.\footnote{We use italic capitals $A$, $B$, \ldots\ to denote {outcomes,} or settings in the case of preparations, and sans-serif capitals \textsf{A}, \textsf{B}, \ldots\ to denote the associated {measurements or preparations.}} 

In the DCES case, the result $C$ of measurement \textsf{C} may be affected by earlier measurement choices $a$ and $b$ at \textsf{A} and \textsf{B} \cite{Mjelva24}. So in the terminology of causal models, the result  of measurement \textsf{C} is a \textit{collider.} 

\begin{figure}[t]
\centering
\begin{tikzpicture}
    \node[draw,rectangle,minimum width=0.6cm,minimum height=0.5cm] (I) at (3.1,-2.5) {\textsf{C}};
\node[] (I0) at (2.5,-3.4) {\textbf{0}};
 \node[] (I1) at (2.9,-3.4) {1};
 \node[] (I2) at (3.3,-3.4) {2};
 \node[] (I3) at (3.7,-3.4) {3};
 \node[] (Prep) at (2.5,-4.7) {\textbf{Preparation}};
 \node[] (C) at (3.1,-4) {$c$};

  \path [dashed,-] (I1) edge (I);
  \path [line width=1.5pt,-] (I0) edge (I);
  \path [dashed,-] (I2) edge (I);
  \path [dashed,-] (I3) edge (I);
  \path [->] (Prep) edge (I0);

    \node[draw,rectangle,minimum width=0.6cm,minimum height=0.5cm] (Abox) at (1.2,0) {\textsf{A}};
   \node[] (Alabel) at (1.2,1.7) {$A$};
    \node[] (A0) at (0.8,1) {0};
    \node[] (A1) at (1.6,1) {1};
    \node[] (a) at (0.5,-1) {$a$};
    \node[draw,rectangle,minimum width=0.6cm,minimum height=0.5cm] (Bbox) at (5,0) {\textsf{B}};
     \node[] (Blabel) at (5,1.7) {$B$};
       \node[] (B0) at (4.6,1) {0};
    \node[] (B1) at (5.4,1) {1};
    \node[] (b) at (5.7,-1) {$b$};

   \path [dashed,-] (Abox) edge (A0);
   \path [dashed,-] (Abox) edge (A1);
   \path [->] (a) edge (Abox.south west);
    \path [dashed,-] (Bbox) edge (B0);
   \path [dashed,-] (Bbox) edge (B1);
    \path [->] (b) edge (Bbox.south east);

   \draw [color=blue,line width=1.5pt,-] (I.north) -- (Abox.south east);
   \draw [color=blue,line width=1.5pt,-] (I.north) -- (Bbox.south west);

\end{tikzpicture}
\caption{The V-shaped case with preparation in state \textbf{0} at input $c$} \label{fig:V}
\end{figure}
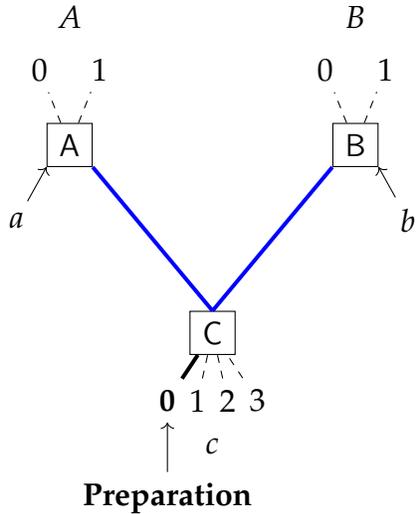

\subsection{Colliders and selection bias}
A {collider} is a variable with more than one direct cause, within a causal model \cite{Tonnies22,Cole10}. In the graphical format of directed acyclic graphs (DAGs), it is a node at which two or more causal arrows converge (hence the term `collider'). 
It is well known that conditioning on such a variable -- i.e., selecting the cases in which it takes a certain value -- may induce a correlation between its  causes, even if they are actually independent. As  \cite{Cole10} puts it, `conditioning on the common effect imparts an association between two otherwise independent variables; we call this selection bias.'\footnote{Collider bias can also hide a correlation between variables that are not independent, as in a  case described by Pigou in 1911 \cite{Pigou11}.}  

Here's an example.\footnote{For which I am indebted to George Davey Smith.} Suppose that Ivy College selects students for academic or athletic excellence, not requiring both, and that these attributes are independent in the general population. 
The admission policy implies that if we learn that an Ivy student is not athletic we can infer that they are academically talented, and vice versa. Within this population, then,  there is a strong anti-correlation between the two attributes. But it is a selection artefact, a manifestation of collider bias. Admission to Ivy is a collider variable, influenced both by athletic ability and academic ability. 

This example illustrates an important point. \textit{Correlations of this kind do not support counterfactuals.} Suppose that Holly is an Ivy track star, who struggles to get As. Would she have had better grades if she'd been less athletic? No. If she had been less athletic she wouldn't have been admitted to Ivy in the first place. (We will say that such correlations are \textit{counterfactually fragile.})

\subsection[Collider bias in DCES versions of W]{Collider bias in DCES versions of W}
Several authors have noted that in the delayed-choice version of the W-shaped experiment, Bell correlations in the postselected ensemble are selection artefacts \cite{Gaasbeek10,Egg13,Fankhauser19,Guido21,PriceWharton21a, Mjelva24}. A mark of this, as we have just seen, is that these correlations need not support the counterfactuals characteristic of Bell nonlocality. A change in Alice's setting $a$ may produce a change in the result $C$ of the intermediate measurement \textsf{C}, thus removing the case from the postselected ensemble. There is hence no need for it make a difference to Bob's outcome $B$. In Bell experiments of this kind, then, the Bell correlations themselves may be counterfactually fragile.

In previous work, Ken Wharton and I called this the \textit{Collider Loophole} for W-shaped tests of nonlocality, and discussed which W-shaped Bell tests are subject to it \cite{PriceWharton21a}. For present purposes, we need only the DCES case, where it is uncontroversial.

 \begin{figure}[t]
\centering
\begin{tikzpicture}
 \node[draw,rectangle,minimum width=0.6cm,minimum height=0.5cm] (I1) at (3,-2.5) {\textsf{S$_1$}};
    \node[draw,rectangle,minimum width=0.6cm,minimum height=0.5cm] (I2) at (7,-2.5) {\textsf{S$_2$}};
    \node[draw,rectangle,minimum width=0.8cm,minimum height=0.5cm] (W) at (5,3) {\textsf{C}};
\node[] (W0) at (4.4,4) {\textbf{0}};
\node[] (W1) at (4.8,4) {1};
 \node[] (W2) at (5.2,4) {2};
 \node[] (W3) at (5.6,4) {3};
  \node[] (Out) at (5,4.6) {$C$};
   \node[] (bc) at (4.4,5.2) {\footnotesize{\textbf{Boundary constraint}}};


 \path [dashed,-] (W1) edge (W);
 \path [line width=1.5pt,-] (W0) edge (W);
\path [dashed,-] (W2) edge (W);
 \path [dashed,-] (W3) edge (W);
\path [->] (bc) edge (W0);

    \node[draw,rectangle,minimum width=0.6cm,minimum height=0.5cm] (Abox) at (1,0) {\textsf{A}};
  \node[] (Alabel) at (1,1.7) {$A$};
    \node[] (A0) at (0.6,1) {0};
    \node[] (A1) at (1.4,1) {1};
    \node[] (a) at (0.3,-1) {$a$};
    \node[draw,rectangle,minimum width=0.6cm,minimum height=0.5cm] (Bbox) at (9,0) {\textsf{B}};
      \node[] (Blabel) at (9,1.7) {$B$};
       \node[] (B0) at (8.6,1) {0};
    \node[] (B1) at (9.4,1) {1};
     \node[] (b) at (9.7,-1) {$b$};

   \path [dashed,-] (Abox) edge (A0);
   \path [dashed,-] (Abox) edge (A1);
    \path [->] (a) edge (Abox.south west);
    \path [dashed,-] (Bbox) edge (B0);
   \path [dashed,-] (Bbox) edge (B1);
   \path [->] (b) edge (Bbox.south east);

      \draw [color=blue,line width=1.5pt,-] (I1.north) -- (Abox.south east);
   \draw [color=blue,line width=1.5pt,-] (I2.north) -- (Bbox.south west);
      \draw [color=blue,line width=1.5pt,-] (I1.north) -- (W.south);
       \draw [color=blue,line width=1.5pt,-] (I2.north) -- (W.south);


\end{tikzpicture}
\caption{The W -shaped case with outcome $0$ at \textsf{C} imposed by a boundary constraint}\label{fig:W2}
\end{figure}
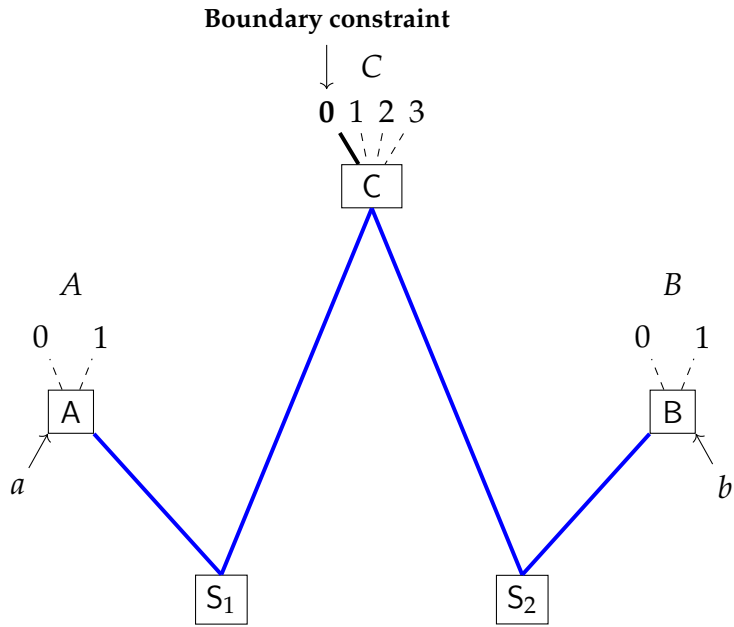

\subsection{Constrained colliders}
This situation changes if the entanglement swapping measurement \textsf{C} is \textit{constrained} to produce a particular outcome, in the way that has been proposed to apply within black holes by Horowitz and Maldacena \cite{HorowitzMaldacena04,Perry21,Perry21b} (see Figure~\ref{fig:W2}). In this special case, the final boundary constraint ensures that a change in Alice's measurement setting cannot make a difference to the result $C$ of the measurement {\textsf{C}}. Bell's Theorem shows it must therefore make a difference to Bob's outcome $B$ (in some cases). The Bell correlation is no longer counterfactually fragile; we shall say it is  \textit{counterfactually robust.} 

Collider constraint is not part of the standard toolbox of causal modelling, but it is easily illustrated by simple examples. Consider Holly, our Ivy College track star. In normal circumstances, as we observed, the anti-correlation between athletic and academic ability does not support counterfactuals. It is not true that if Holly had not been athletic, she would have been academically talented. But now imagine that Fate has taken an interest in Holly's career, and has decreed since her birth that she will be admitted to Ivy. In this case, Fate has constrained the collider variable, fixing it to the positive value (`admitted'). Given the admission policy, it is now \textit{true} that if Holly had not been athletic she would have been academically talented. The counterfactual is now robust, not fragile. 
    
Returning to the real world, the Horowitz-Maldacena proposal does for measurement inside a black hole what Fate does for Holly. It constrains the final measurement outcome to have a particular value. Maldacena and Horowitz suggest that this
creates a zigzag causal path through time, along which information can
escape from a black hole.\footnote{For recent discussion see \cite{Perry21,Perry21b}.} It is important that the process supports counterfactuals. If different information had fallen into the black hole in the first place, different information would have emerged. 

In this very special case, we have shown how robust counterfactual-supporting Bell nonlocality can result from a combination of two things: (i) collider bias, and (ii) a boundary constraint, imposing a particular value on a collider variable. Elsewhere, Ken Wharton and I have called this kind of counterfactual-supporting relation \textit{Connection across a Constrained Collider} (CCC) \cite{PriceWharton21b,PriceWharton23}. 

The remainder of the present argument aims to show that the same explanation works in a regular V-shaped Bell experiment (Figure~\ref{fig:V}), with ordinary measurement preparation providing the required boundary constraint.

\subsection[Application to the V-shaped case?]{Application to the V-shaped case?}
 The obvious obstacle to applying the same story to the V-shaped case is that the input variable at \textsf{C} in Figure~\ref{fig:V} does not seem to be a collider. In the familiar case, in which the measurement is simply \textit{prepared} in a particular initial state at \textsf{C}, that initial state is obviously not influenced by the measurement settings at \textsf{A} and \textsf{B}. 

However,  this is not the relevant case to consider. After all, the final state at \textsf{C} isn't affected by Alice and Bob's choice of measurement settings, in the case in which it is constrained, as in Figure~\ref{fig:W2}. In that case, the outcome at \textsf{C} is fixed by the boundary constraint. What matters is that it's a collider \textit{when not constrained.}

So in the V-shaped case, too, we need to consider the unconstrained case. Unlike in the W-shaped case, however, the unconstrained V-shaped case is highly unusual, and can easily escape notice altogether. We are going to propose two ways to bring it into view. One of them is \textit{local,} in the sense that it focuses simply on experiments like that shown in Figure~\ref{fig:V}, and proposes a technique to make the initial variable unconstrained. The other is \textit{global,} in that it asks us to consider whatever time-asymmetric feature of reality normally gives us control of the initial but not the final conditions of experiments, and consider an idealised regime in which that feature is absent. (We will call this the \textit{constraint-free regime}.)

\begin{figure}[t]
\centering
\begin{tikzpicture}
    \node[draw,rectangle,minimum width=0.6cm,minimum height=0.5cm] (I) at (3.1,-2.5) {{\textsf{C}}};
\node[] (I0) at (2.5,-3.4) {0};
 \node[] (I1) at (2.9,-3.4) {1};
 \node[] (I2) at (3.3,-3.4) {2};
 \node[] (I3) at (3.7,-3.4) {3};
 \node[] (C) at (3.1,-4) {$c$};

 \path [dashed,-] (I1) edge (I);
  \path [dashed,-] (I0) edge (I);
  \path [dashed,-] (I2) edge (I);
  \path [dashed,-] (I3) edge (I);

 \node[] (D0) at (2.5,3.7) {0};
 \node[] (D1) at (2.9,3.7) {1};
 \node[] (D2) at (3.3,3.7) {2};
 \node[] (D3) at (3.7,3.7) {3};
 \node[] (Dout) at (3.1,4.3) {$D$};

     \node[draw,rectangle,minimum width=0.6cm,minimum height=0.5cm] (D) at (3.1,2.8) {\textsf{D}};

\path [dashed,-] (D1) edge (D);
  \path [dashed,-] (D0) edge (D);
 \path [dashed,-] (D2) edge (D);
 \path [dashed,-] (D3) edge (D);

    \node[draw,rectangle,minimum width=0.6cm,minimum height=0.5cm] (Abox) at (1.2,0) {{\textsf{A}}};
   \node[] (Alabel) at (1.2,1.7) {$A$};
    \node[] (A0) at (0.8,1) {0};
    \node[] (A1) at (1.6,1) {1};
    \node[] (a) at (0.5,-1) {$a$};
    \node[draw,rectangle,minimum width=0.6cm,minimum height=0.5cm] (Bbox) at (5,0) {{\textsf{B}}};
    \node[] (Blabel) at (5,1.7) {$B$};
       \node[] (B0) at (4.6,1) {0};
    \node[] (B1) at (5.4,1) {1};
    \node[] (b) at (5.7,-1) {$b$};

   \path [dashed,-] (Abox) edge (A0);
   \path [dashed,-] (Abox) edge (A1);
   \path [->] (a) edge (Abox.south west);
    \path [dashed,-] (Bbox) edge (B0);
   \path [dashed,-] (Bbox) edge (B1);
    \path [->] (b) edge (Bbox.south east);

   \draw [color=blue,line width=1.5pt,-] (I.north) -- (Abox.south east);
   \draw [color=blue,line width=1.5pt,-] (I.north) -- (Bbox.south west);


    \draw[color=red,line width=1pt] (a) to (Abox.south) to (I.north west);
    \draw[color=red,line width=1pt] (b) to (Bbox.south) to (I.north east);


     \draw[color=gray,dashed,rounded corners,line width=1pt] (I) -- (D);

\end{tikzpicture}
\caption{V-shaped case with delayed-choice measurement of initial state} \label{fig:V2a}
\end{figure}
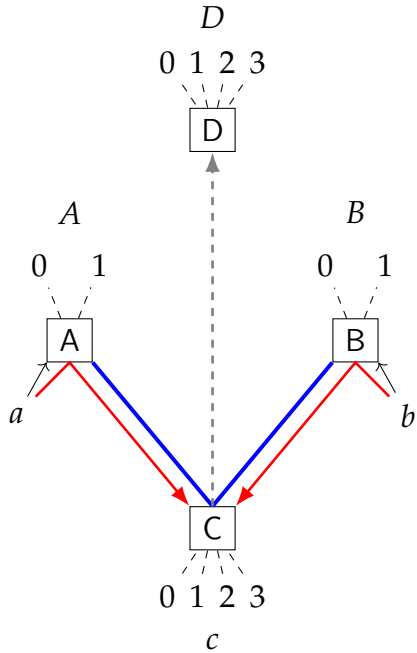

\subsubsection[A delayed-choice V-shaped experiment?]{A delayed-choice V-shaped experiment?}

For the local argument, imagine the version of the V-shaped experiment shown in Figure~\ref{fig:V2a}, in which the four possible initial Bell states are produced at \textsf{C}, with equal probability. In each run of the experiment the initial state is recorded, but by a device that does not produce a classical record until a measurement \textsf{D}, in the absolute future of the measurements at \textsf{A} and \textsf{B}. (This is the delayed-choice element.)\footnote{I am indebted to Gerard Milburn for this proposal.}

The expected statistics for this experiment are straightforward. There are no Bell correlations in the statistics as a whole, but such correlations emerge if we postselect on any of the four possible results of \textsf{D} (or, equivalently, preselect on any of the four initial states at \textsf{C}).

How should we interpret these results? Are the Bell correlations in these postselected ensembles counterfactually \textit{fragile,} as in Figure~\ref{fig:W}, or counterfactually \textit{robust,} as in Figure~\ref{fig:W2}? The question turns on whether the settings choices at \textsf{A} and \textsf{B} are able to influence the results of the measurement at \textsf{D}. If such influence is possible, then the Bell correlations in each subensemble can be explained as a selection artefact -- no counterfactually robust nonlocality is required. 

How should we answer this question? Indeed, does it have a settled answer, one way or the other, or is the case under-described? Perhaps there are different possible versions of such an experiment, some going one way and some the other. 

For our purposes, it isn't necessary to resolve these issues. For us, the case is a thought experiment. If the reader will allow the possibility of versions of this experiment in which the Bell correlations in each subensemble are a selection artefact, then we have the model we need. We have a V-shaped case in which the initial state is unconstrained, from which the familiar V-shaped case in Figure~\ref{fig:V} may then be obtained by reimposing the usual choice of the initial state.

Note, however, that if the setting choices at \textsf{A} and \textsf{B} can make a difference to the outcome at \textsf{D}, then they are also making a difference at \textsf{C}. The case then involves \textit{retrocausality,} as shown in red in Figure~\ref{fig:V2a}. 
Some readers will balk at this possibility. Invoking the words of John Wheeler, perhaps, they may object that it allows that `present choice influences past dynamics, in contravention of every principle of causality' \cite[p.~41]{Wheeler78}.
It would take us too far afield to challenge Wheeler's instincts on this matter.\footnote{Ken Wharton and I have done so at length elsewhere \cite{Price96,PriceWharton15,PriceWharton16,WhartonArgaman20}.   In the present context, we may be overstating the relevance of Wheeler's objection. Wheeler is discussing a delayed-choice experiment, and the full passage runs like this:
\begin{quote}
    Does this result mean the present choice influences past dynamics, in contravention of every principle of causality? Or does it mean, calculate pedantically and don’t ask questions? Neither; the lesson presents itself rather as this, that the past has no existence except as it is recorded in the present.
\end{quote}
Wheeler's own option may well be sufficient for our present purposes. If in Figure~\ref{fig:V2a} the measurement settings at \textsf{A} and \textsf{B} can make a difference to the past recorded at \textsf{D}, that seems good enough.} But for readers wedded to those instincts, we need a more powerful intuition pump. 

\subsubsection{The constraint-free regime}
 As already noted, ordinary experimental control is time-asymmetric. We control the \textit{initial} conditions of experiments, but not their \textit{final} conditions. Let's call this time-asymmetry \textit{Initial Control.} Some writers call it simply \textit{Causality} \cite{Chiribella10,Coecke14}. Plausibly, it is a result of the thermodynamic asymmetry, though this isn't essential to our argument. (For discussion,  see \cite{Price96, Price10,Albert02,PriceWeslake10,Rovelli21}.) 
 Whatever its source, we can consider an imaginary regime in which it is absent. This is our {{constraint-free regime (CFR).}} It is comparable to the ideal frictionless regime in mechanics (friction being similarly time-asymmetric, of course).

\begin{figure}[t]
\centering
\begin{tikzpicture}
    \node[draw,rectangle,minimum width=0.6cm,minimum height=0.5cm] (I) at (3.1,-2.5) {\textsf{C}};
\node[] (I0) at (2.5,-3.4) {0};
 \node[] (I1) at (2.9,-3.4) {1};
 \node[] (I2) at (3.3,-3.4) {2};
 \node[] (I3) at (3.7,-3.4) {3};
  \node[] (C) at (3.1,-4) {$c$};

  \path [dashed,-] (I1) edge (I);
  \path [dashed,-] (I0) edge (I);
  \path [dashed,-] (I2) edge (I);
  \path [dashed,-] (I3) edge (I);

    \node[draw,rectangle,minimum width=0.6cm,minimum height=0.5cm] (Abox) at (1.2,0) {\textsf{A}};
    \node[] (Alabel) at (1.2,1.7) {$A$};
    \node[] (A0) at (0.8,1) {0};
    \node[] (A1) at (1.6,1) {1};
    \node[] (a) at (0.5,-1) {$a$};
    \node[draw,rectangle,minimum width=0.6cm,minimum height=0.5cm] (Bbox) at (5,0) {\textsf{B}};
      \node[] (Blabel) at (5,1.7) {$B$};
       \node[] (B0) at (4.6,1) {0};
    \node[] (B1) at (5.4,1) {1};
    \node[] (b) at (5.7,-1) {$b$};

   \path [dashed,-] (Abox) edge (A0);
   \path [dashed,-] (Abox) edge (A1);
   \path [->] (a) edge (Abox.south west);
    \path [dashed,-] (Bbox) edge (B0);
   \path [dashed,-] (Bbox) edge (B1);
    \path [->] (b) edge (Bbox.south east);

   \draw [color=blue,line width=1.5pt,-] (I.north) -- (Abox.south east);
   \draw [color=blue,line width=1.5pt,-] (I.north) -- (Bbox.south west);


    \draw[color=olive,line width=1pt] (a) to (Abox.south) to (I.north west);
    \draw[color=olive,line width=1pt] (b) to (Bbox.south) to (I.north east);

\end{tikzpicture}
\caption{V-shaped case in CFR} \label{fig:V2}
\end{figure}
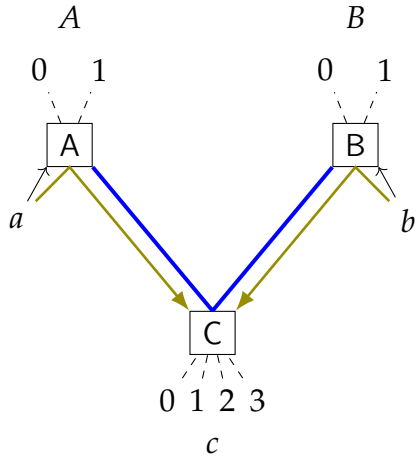

In CFR, causation and control are by definition time-symmetric, \textit{to the extent that they survive at all.} The qualification is important. If abandoning Initial Control meant abandoning causality altogether, we could no longer speak of colliders, in either temporal direction. 

If nothing else, however, CFR will preserve the two-way determination by boundary conditions that Hawking \cite[p.~346]{Hawking94} has in mind, when he declares that there is no such thing as a time-asymmetry of causation.\footnote{As philosophers have wrongly supposed, in Hawking's view. He criticises Reichenbach in particular. As we have just seen, Wheeler is equally insistent that causation \textit{is} time-asymmetric. The most likely diagnosis of the disagreement is that Hawking and Wheeler are talking about different things, but we needn't explore this here.} 

\begin{quote}
[I]n physics we believe that there are laws that determine the evolution of the universe uniquely. So if state A evolved into state B, one could say that A caused B. But one could equally well look at it in the other direction of time, and say that B caused A. So causality does not define a direction of time.  
\end{quote}
With (a probabilistic version of) this weak notion as a fallback, 
we can say that CFR permits \textit{some} notion of causality. By considering hypothetical changes to conditions on boundaries, we can also save a weak counterfactual notion of influence, or making a difference. And it will all be time-symmetric, just as Hawking says.

In CFR, then -- with these caveats about what we mean -- we have no reason to exclude the possibility that settings choices $a$ and $b$ `influence' the initial state $c$ at \textsf{C}, as shown in olive green in Figure~\ref{fig:V2}. (We have changed the colour of the arrows to remind readers that causality means something different than in Figure~\ref{fig:V2a}.) The initial state thus becomes a collider, and Bell correlations can now be explained as collider artefacts, just as in the delayed-choice W-shaped case.\footnote{Some readers may be troubled by the presumption that there could be experiments at all in CFR. But here we are on well-trodden ground. A community familiar with  Boltzmann Brains  should not be troubled by a few simple pieces of quantum optics. Such things will arise by chance in a sufficiently expansive CFR, and we may discuss their characteristic statistics.} 

Specifically, there is no need for a change in Alice's measurement setting $a$ ever to make a difference to Bob's outcome $B$, because the difference can always be absorbed as a change in the input state  $c$. Again, this is exactly as in the delayed-choice W case, in its normal version, without a boundary constraint at the collider.

 \subsection{Reimposing Initial Control}

 Whether we take the local or the global path to an unconstrained version of the V-shaped experiment, the normal case in Figure~\ref{fig:V} is exactly what we'd expect, if we reimpose Initial Control at \textsf{C}. So the proposed explanation of the Bell correlations in Figure~\ref{fig:V} works just as well here as in the delayed-choice W case with the Horowitz-Maldecena boundary constraint (Figure~\ref{fig:W2}). 
 
 We suspect that the reason this parallel hasn't been noticed is that what's normal in the two cases is completely reversed. For W, lack of a boundary constraint is normal, and the constrained case is exceptional. For V, lack of boundary constraint is exceptional, and the constrained case is normal. The logic is exactly the same in both experiments, but to see the parallels we need to put the two exceptional cases on the table, as well as the two normal cases. The table then looks as follows. The greyed-out options are those that are `hard to see', not being manifest in ordinary circumstances.

\begin{table}[ht]
\begin{center}
\resizebox{\textwidth}{!}{%
\begin{tabular}{@{}rll@{}}
\toprule
\multicolumn{1}{l}{}                   & Unconstrained colliders               & Constrained colliders                                     \\ \midrule
\multicolumn{1}{r}{W} & \multicolumn{1}{l}{Normal} & \multicolumn{1}{l}{\textcolor{gray}{Exceptional (only  in black holes?)}} \\ \cmidrule(l){2-3} 
V                      & \textcolor{gray}{Exceptional (only in CFR?)}                 & Normal                                          \\ \bottomrule
\end{tabular}%
}
\end{center}
\end{table}
\ \\\vspace{-48pt}

 \subsection{Conclusion}
 We have argued that in both the V and W-shaped cases, Bell nonlocality between \textsf{A} and \textsf{B} can result from a combination of two things: (i) causal influences operating within the lightcones; and (ii) boundary constraints, restricting the possible values of collider variables. This turns out to be capable of  producing a counterfactual-supporting spacelike influence, even though neither of the individual ingredients is in tension with special relativity. It has long been recognised that in principle, zigzag influences might render EPR phenomena compatible with relativity \cite{Costa53,Price96,PriceWharton15,WhartonArgaman20}. This proposal offers a mechanism to produce such zigzag influences. 

 It may be helpful to frame our argument as lying at the intersection of  two familiar sources of illumination. On one side, it is standard fare of causal modelling that conditioning on colliders produces selection artefacts. If we turn this spotlight on QM, it turns out to be uncontroversial that \textit{some} cases of Bell correlations (i.e., those in DCES W-shaped experiments) are properly explained this way.\footnote{Note that it is trivial to show that postselection can produce Bell correlations in toy models \cite{PriceWharton23}. The interesting fact is that it actually happens at a fundamental level in QM.}

 From the other side, it is a familiar proposal that most of the physical manifestations of `time's arrow' -- in other words, most of the observed time-asymmetry in our surroundings -- results from imposing a low-entropy boundary condition on a time-symmetric space of possible histories. This is often called `conditioning on the Past Hypothesis (PH)' \cite{Albert02}. Note that `conditioning' means something a little different in these two cases. No one supposes that to condition on PH, we have to ignore a vast number of \textit{actual} universes in which PH does not hold. But it is the same mathematical idea. 

 Turning this second spotlight on QM, we have proposed that ordinary V-shaped Bell correlations can {also} be understood as selection artefacts -- a result of conditioning on their initial conditions, in the familiar way permitted by Initial Control. Plausibly, this is another of the local, downstream manifestations of the low entropy past. 

 Which notion of conditioning does our proposal involve? It is closer to the second, in the sense that the proposal does not require that Initial Control enables us to select from a range of \textit{actual} alternatives. We mention this in case readers are bothered by this difference between pre- and post-conditioning. There is indeed a difference, but the notion of pre-conditioning we need is already in play, apparently doing important work in explaining time-asymmetry.  Our proposal is simply a new application of that familiar idea, in the context of the correlations permitted by QM.
 
  \subsection{An objection}

  We close by mentioning an objection. It might be pointed out that the usual causal story in the delayed-choice W case (Figures~\ref{fig:W} and \ref{fig:W2}) is \textit{not} confined to the lightcones: it involves a direct `instantaneous' projection on the inner particles, within each wing of the experiment, when measurements are performed at \textsf{A} and \textsf{B}. This objection is true as far as it goes, but it does not have any bearing on the ability of the proposal to provide a relativity-friendly explanation of nonlocality in the V-shaped case. Moreover, it seems reasonable to hope that the same kind of account will avoid the need for such instantaneous projection in the V-shaped wings of the W experiment, once implemented at the level of a suitable ontology.\footnote{Like everything I have written on QM for at least a decade and a half, this piece is greatly indebted to conversations with Ken Wharton. I am also grateful to Gerard Milburn, Michael Cuffaro, Jeff Bub, Heinrich P\"as, for discussion and helpful comments on related material.}


\end{document}